\documentclass[preprint,showpacs,showkeys]{revtex4}
\usepackage{epsf, epsfig, latexsym}
\usepackage{graphicx}
\usepackage{amssymb}
\usepackage{amsmath}
\usepackage{bm}
\usepackage{bbm}
\usepackage{dcolumn}
\usepackage{multirow}
\usepackage{pifont}
\usepackage{calc}%
\begin{document}
\bibliographystyle{apsrev}
\title[]
{Gluon Mass from Curvature of Gauge Slice}
\author{Eun-Joo \surname{Kim}}
\email[]{ejkim@jbnu.ac.kr}
\author{Jong Bum \surname{Choi}}
\email[]{jbchoi@jbnu.ac.kr}
\affiliation{Division of Science Education and Institute of Science Education,
Chonbuk National University, Jeonju 561-756, Korea}
%
%
\date{\today}
%
%
\begin{abstract}
%
The masslessness of gluons due to gauge invariance appears to be contradictory
with the observed massive structures of hadrons. As a dual process of gauge transformation,
we have to fix the gauge to quantize the gluons in order to get propagators.
The contradictory picture between gauge invariance and massive gluon propagation
with vacuum condensates can be overcome by introducing generalized $\theta$ vacuum
composed of curved gauge slices. On the curved gauge slices the gluon propagator
turns out to be dependent on the curvature and in maximally symmetric space
the curvature plays the role of mass, thus providing a solution to the
contradiction between masslessness and massive propagation.
\end{abstract}
%
%

\maketitle

%
%
The rest mass of a particle is usually measured by comparing momentum and
velocity, but in the case of gauge bosons the velocity is either fixed to $c$ or
cannot be measured because of very short life times. The masses of weak gauge
bosons $W^{\pm}$ and $Z^0$ are deduced from the initial or final four momenta of the interacting
particles and therefore represent just the resonant energy states which are mediating
weak interactions. Since the range of propagation is very short, the contradictory aspects
between the massiveness of resonance and the masslessness as gauge bosons can be
accommodated by introducing vacuum filled with condensates spawned by assumed Higgs
field. Now the Higgs field has been confirmed experimentally and vacuum condensates
have become a reality to explain the massiveness of $W^{\pm}$ and $Z^0$.

On the other hand, vacuum condensates in quantum chromodynamics (QCD) originate from
the nonperturbative interactions of gluons. These gluons are taken to be massless
because they are introduced as gauge bosons and there still remains the problem of
reconciliation of masslessness with the observed hadronic masses. In order to account for
hadronic masses it is convenient to introduce gluonic mass, but the inclusion of mass
term is not consistent with the gauge invariance principle. Therefore we need to
devise a new method to combine gauge invariance and gluonic mass effect.

The first argument of massive gauge boson was based on the possibility of removal
of the weak coupling assumption. With strong vector coupling, gauge invariance could be
restored with the aid of vacuum expectation values of time-ordered operator products~\cite{R1}.
This kind of Schwinger mechanism had been shown to be accomplished through dynamical
symmetry breaking via self-interactions in the case of massless Yang-Mills theory~\cite{R2}.
However, the self-interactions are nonperturbative and the dynamically generated gluon
mass turns out to be dependent on momentum resulting in vanishment at short distances~\cite{R3}.
In this case the QCD vacuum is taken to be a tangle of fluctuating color fields
and the effects of fluctuations can be parametrized by introducing various vacuum condensates~\cite{R4}.
These vacuum condensates can change the forms of quark and gluon propagators and there
appears naturally the idea of running quark mass~\cite{R5}.
But the gluon propagator remains in massless form
when we consider only gauge invariant operators to define vacuum condensates.
A turning point has been made by considering dimension-2 condensates.

Originally dimension-2 condensate was considered for systematic operator product expansion of
QCD propagators~\cite{R6} and ghost fields were used to
account for the dynamical generation of gluonic mass term~\cite{R7}.
Then it seemed quite natural to extend the field operators into the gluonic ones~\cite{R8}.
However, in this case, the condensate value depends on the choice of gauge
and these dependences can be used to define gauge slices.
In flat field space, gauge slices are taken to be in the form of hyperplane when
the gauge condition is given by linear functional of gauge field $\mathbf{A}_{\mu}$.
This situation changes if we consider nonperturbative interactions of gluons.
Because of nonperturbative self-interactions we can induce that the gluonic states
cannot be classified according to the number of gluons. Logically the unique possibility
to define the gluonic state is to consider the state composed of infinite number of gluons.
Let this state be represented by $|\Omega \rangle$ and then we have
\begin{equation}
   A_{\mu}^{a}(x) |\Omega \rangle  =  \beta |\Omega \rangle
\label{eq1}
\end{equation}
because the gluon number is not changed by one creation or by one annihilation. Inner
production of each side of this equation results in the covariant relation
\begin{equation}
   \langle \Omega| A_{\mu}^{a}(x)A_{a}^{\mu}(x) |\Omega \rangle  =  \alpha^2.
\label{eq2}
\end{equation}
This relation defines the dimension-2 condensate and the state $|\Omega \rangle$
can be taken to represent a condensed vacuum. For different $\alpha$ values we can get
different vacua and this situation is quite similar to the case of $\theta$ vacua.
The $\theta$ vacuum is defined by linear combination of local vacuum states deduced from the
instanton solution and the eigenvalue of gauge transformation is represented by the value of
$\theta$. Although the eigenstate is fixed by the eigenvalue $e^{-i\theta}$,
the angle $\theta$ remains as a free parameter representing different vacuum states
physically independent from each other~\cite{R9}.
The relation between the states $|\Omega \rangle$ and $\theta$ vacua
can be made into a one-to-one correspondence if we combine the ideas of
in-hadron condensates~\cite{R10} and generalization of $\theta$ vacuum with spacetime dependent
$\theta(\mathbf{x}, t)$~\cite{R11}. The spacetime dependent $\theta$ has been introduced to account
for parity-odd domains formed in relativistic heavy-ion collisions~\cite{R12}.

In-hadron condensates have been introduced in the analysis of bound states in QCD.
Because of color confinement, quark and gluon condensates can be associated
with the dynamics of hadronic wavefunctions rather than the vacuum itself~\cite{R13}.
These notions can be used to cure the problem of huge cosmological constant
resulting from conventional condensates associated with ordinary vacuum~\cite{R14}.
But there appears the problem of hadronic boundary if we consider spacetime constant condensate.
The other choice is to introduce spacetime varying condensates inside hadrons
and the variations can be represented with different $\alpha$ values in Eq.~(\ref{eq2}).
A natural solution to the problem of hadronic boundary is to assume smooth variations of
$\alpha$ into zero for the outer region of a hadron. Then for a fixed value of $\alpha$
the set of points $x$ satisfying the condition in Eq.~(\ref{eq2}) are expected to form
a closed surface inside a hadron. For different $\alpha$ values the closed surfaces form
a set of slices which can be taken to represent each gauge fixed vacuum state $|\Omega \rangle$.

On the other hand, the generalization of $\theta$ vacuum with spacetime dependent $\theta$
has been proposed to account for the possible parity violation in an excited vacuum domain formed
in relativistic heavy-ion collisions. The spacetime dependence of $\theta$ can be transferred into
complex mass parameter and the effects of this complex mass parameter turn out to be the creation
of quark-antiquark pairs with nonzero chirality. These nonzero chiral pairs can generate spatial
asymmetry in the production of charged pions~\cite{R15}
and the asymmetries have been observed in heavy-ion experiments~\cite{R16}.
The observations of asymmetries are interpreted as the evidence of the formation of
a domain with $\theta = \theta(\mathbf{x}, t)$ that can be assigned to new phase of quark-gluon matter.
Since the formed domain is restricted to the collided region of heavy ions, the structure of
the points with constant $\theta$ is expected to be a closed surface inside the collided region.
For different $\theta$ values the closed surfaces form a set of slices which represent physically
independent $\theta$ vacua.

What is the relation between vacuum states $|\Omega \rangle$ and $\theta$ vacua?
The state $|\Omega \rangle$ is defined by the value of dimension-2 condensate and $\theta$ vacua
have been introduced by defining local vacuum states deduced from instanton solution.
If we can establish some relations between instantons and dimension-2 condensates,
it may be possible to unify the definitions of vacuum for strongly interacting systems.
Fortunately there have been attempts to prove the origin of dimension-2 condensate
as the contributions of instantons~\cite{R17}.
The method of proof had been restricted to lattice calculations due to the nonperturbative
nature of the problem. By employing different methods to cross check the results,
it has been possible to prove that the dimension-2 condensates receive significant instantonic
contributions. The slight differences are due to the fact that dimension-2 condensates are mainly
defined inside hadrons but instanton solutions exist over some extended region.
The unification of definitions of QCD vacuum can be achieved if we take the collections of curved
sets of slices defined by Eq.~(\ref{eq2}) as representing generalized $\theta$ vacuum~\cite{R18}.
For each $\theta$ value only one $\alpha$ value is assigned and this one-to-one correspondence
can be exploited to extend the definition of QCD vacuum. Now the gauge slices can be drawn
from the inner region of a hadron to the outer region of heavy ion or quark-gluon matter.

\begin{figure}[htb]
 \includegraphics[width=0.76\linewidth, angle=-86]{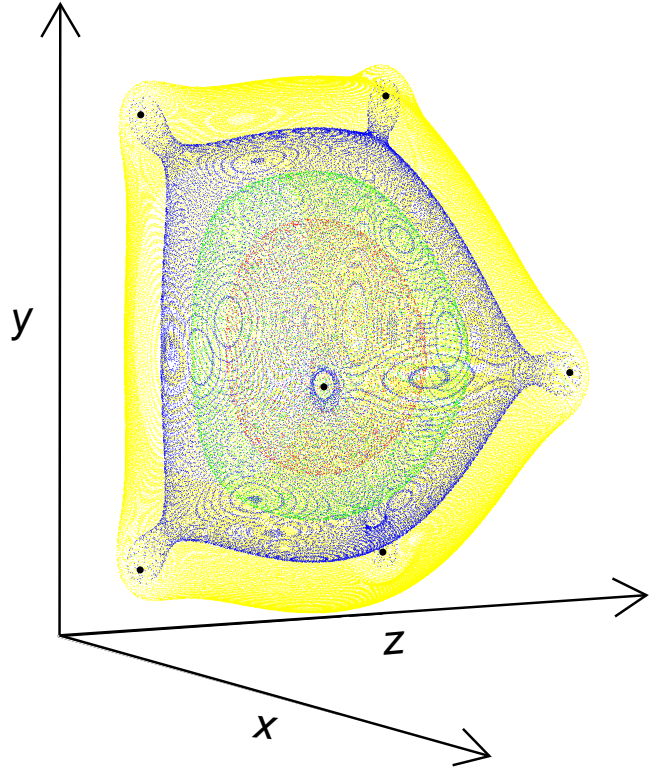}	
\caption{(Color online)
 Equi-$\langle A_{\mu}^{2} \rangle$ surfaces for a deuteron with quarks
      at $(-\frac{\sqrt{3}}{4}, -0.75, 0.75)$, $(\frac{\sqrt{3}}{2}, 0.0, 0.75)$,$(-\frac{\sqrt{3}}{4}, 0.75, 0.75)$,
      $(-\frac{\sqrt{3}}{4}, -0.75, -0.75)$, $(\frac{\sqrt{3}}{2}, 0.0, -0.75)$ and $(-\frac{\sqrt{3}}{4}, 0.75, -0.75)$.}
 \label{fig1}
 \end{figure}

The main transition of viewpoint with new QCD vacuum from the ordinary one is the curvedness of
each gauge slice. This transition is essential for in-hadron condensates and naturally extended
to generalized $\theta$ vacuum. In order to deduce the form of curvedness we need to introduce
necessary assumptions and have to construct appropriate model based on these assumptions.
The assumptions are related to the change of QCD vacuum state according to creation or
annihilation of quark pairs. These changes can be described systematically by classifying
vacuum domains with given numbers of quarks and antiquarks~\cite{R19},
and for classified domains appropriate measures can be defined to formulate the curvedness.
One example is the amplitude defined by nonlocal quark condensate~\cite{R20}
 \begin{equation}
   \langle : \bar{q}(x)q(0) : \rangle \equiv \langle : \bar{q}(0)q(0) : \rangle Q(x),
\label{eq3}
\end{equation}
where the form of $Q(x)$ can be given in one model as~\cite{R21}
 \begin{equation}
   Q(x) = \frac{Q_0}{x^\beta} \exp \left \{ -\frac{1}{k} \frac{x^2 - x }{\ln |x|} \right\}
\label{eq4}
\end{equation}
with $x$ normalized by appropriate scale parameter. Then by assuming the value of
dimension-2 condensate to be proportional to the probability amplitude to have
a quark pair at that point, we get for a 6-quark domain representing a deuteron~\cite{R22}
 \begin{equation}
   \langle A_{\mu}^{2} \rangle = A_0^2 \prod_{i=1}^{6} Q_{xi}
      \left \{ \sum_{i=1}^{6} \prod_{r_j, r_k \neq r_i}Q_{jk}
              + \sum_{r_j, r_k} Q_{jk} \cdot \prod_{r_\alpha, r_\gamma \neq r_j, r_k} Q_{\alpha \gamma}  \right\}
\label{eq5}
\end{equation}
with the notation
 \begin{equation}
   Q_{ij} = \frac{Q(r_{ij})}{Q_0} = \frac{Q(|\mathbf{r}_i - \mathbf{r}_j |)}{Q_0}.
\label{eq6}
\end{equation}
Several gauge slices are drawn in Fig.~\ref{fig1} with the formula given in Eq.~(\ref{eq5}).
The positions of the 6 quarks are chosen such that they stay at the corners of a trigonal prism.
%

Now the quantization of gluonic field has to be carried out on these curved gauge slices.
To account for the curvedness we need to replace the ordinary derivative by the covariant
derivative defined as
\begin{equation}
  D_{\mu}\mathbf{A}_{\nu} \equiv \frac{\partial\mathbf{A}_{\nu}}{\partial x^{\mu}} - \Gamma_{\nu\mu}^{\kappa}\mathbf{A}_{\kappa}  ,
\label{eq7}
\end{equation}
where $\Gamma_{\nu\mu}^{\kappa}$ represents the affine connection
which relates the ordinary coordinates to the freely falling coordinates.
The free Lagrangian for gluon propagator on curved gauge slice is given by
\begin{equation}
  L = -\frac{1}{4} \int d^4 x   ~(D_{\mu}\mathbf{A}_{\nu} - D_{\nu}\mathbf{A}_{\mu} )
   \cdot ( D^{\mu}\mathbf{A}^{\nu} - D^{\nu}\mathbf{A}^{\mu}  )  .
\label{eq8}
\end{equation}
To carry out functional integration we have to arrange the differential operators into the form
\begin{equation}
    L = \frac{1}{2} \int d^4 x ~\mathbf{A}_{\mu} \cdot  ( D^2 g^{\mu\nu} - D^{\nu}D^{\mu}  ) \mathbf{A}_{\nu}  ,
\label{eq9}
\end{equation}
where the covariant differential operators satisfy the commutation relation~\cite{R23}
\begin{equation}
    D_{\nu}D_{\mu}\mathbf{A}_{\kappa} - D_{\mu}D_{\nu}\mathbf{A}_{\kappa}
      = -\mathbf{A}_{\sigma}R_{~\kappa\mu\nu}^{\sigma}  .
\label{eq10}
\end{equation}
The Riemann-Christoffel curvature tensor $R_{~\kappa\mu\nu}^{\sigma}$ is defined as
\begin{equation}
    R_{~\kappa\mu\nu}^{\sigma}  \equiv
       \frac{\partial\Gamma_{\kappa\mu}^{\sigma}}{\partial x^{\nu}} -
       \frac{\partial\Gamma_{\kappa\nu}^{\sigma}}{\partial x^{\mu}} +
       \Gamma_{\kappa\mu}^{\eta}\Gamma_{\nu\eta}^{\sigma} -
       \Gamma_{\kappa\nu}^{\eta}\Gamma_{\mu\eta}^{\sigma}  ,
\label{eq11}
\end{equation}
and the contracted Ricci tensor is given by
\begin{equation}
    R_{\mu\nu} = R_{~\mu\sigma\nu}^{\sigma}  .
\label{eq12}
\end{equation}
Then we can rearrange the covariant differential operators and get
\begin{equation}
    L = \frac{1}{2} \int d^4 x ~\mathbf{A}_{\mu} \cdot  ( D^2 g^{\mu\nu} - D^{\mu}D^{\nu} + R^{\mu\nu} ) \mathbf{A}_{\nu}
\label{eq13}
\end{equation}
with additional Ricci tensor $R^{\mu\nu}$. The gluon propagator is obtained as the inverse matrix of the tensor
that appears between the two gluonic fields. In general $R^{\mu\nu}$ varies from point to point and
it is non-trivial to deduce the form of gluon propagator.
In case of flat space with vanishing Ricci tensor
we can get the ordinary form of gluon propagator without any mass parameter.
The next simple case is that of maximally symmetric gauge slice where we can write~\cite{R24}
\begin{equation}
    R^{\mu\nu} = m^{2}g^{\mu\nu} .
\label{eq14}
\end{equation}
The parameter $m^2$ is related to the curvature scalar and the gluon propagator can be obtained as
\begin{equation}
    D_{ab}^{\mu\nu}(k) = \frac{1}{k^2 - m^2}\Big ( g^{\mu\nu} - \frac{ k^{\mu}k^{\nu} }{k^2} \Big ) \delta_{ab} ,
\label{eq15}
\end{equation}
which is just the propagator form of a particle with momentum $k$ and mass $m$.
Thus gluon seems to get a mass by propagating on curved gauge slice.
For general $R^{\mu\nu}$ it is possible to have a minus sign in Eq.~(\ref{eq14})
resulting in the idea of imaginary mass.

The propagation of gluons through curved gauge slices can give solutions to various problems.
First of all, the differences between current quark mass and constituent quark mass and
even dynamical quark mass can be explained systematically.
The size of the extended gluonic domain is related to the estimation of quark mass.
Furthermore these extended gluonic domains can be used to describe multiquark states
such as tetraquarks and are expected to be applicable to the descriptions of exotic nuclei
such as $^{11}\text{Li}$~\cite{R25} and $4\alpha$ linear chain configuration of $^{16}\text{O}$~\cite{R26}.
For heavy ion collisions, particle production processes from the new phase of quark-gluon matter
are related to the structures of gluonic domains formed by a set of curved slices
which represent generalized $\theta$ vacua. This generalized picture is useful to account for the
observed parity violation in hot quark-gluon matter state.
Macroscopically, gluonic domains can be defined for strongly interacting objects like neutron stars.
Then it is possible to extend the domains into the outside region of neutron star
for very low energy gluons~\cite{R27}. These gluons can generate mass effects
without any interactions with charged particles or photons.
Thus low energy gluons can behave as dark matter originating from the initial state of the Big Bang.
Since the mass parameter is related to the curvature of large scale gauge slice as in Eq.~(\ref{eq14}),
the $\frac{1}{r^2}$ variation of dark matter density in a galaxy can be easily explained.
Moreover, for inter-galactic regions, imaginary mass effects from Eq.~(\ref{eq14}) may result
in repulsive gravity providing for another solution of the problem of cosmic acceleration~\cite{R28}.

In summary, we have tried to combine the ideas of in-hadron condensates and generalized
$\theta$ vacuum resulting in new QCD vacuum composed of curved gauge slices.
The gluonic fields have to be quantized on these curved gauge slices and the curvedness
can be accounted as covariant derivatives in the Lagrangian.
Compared with flat space calculations, there appears additional Ricci tensor
and the effect of this additional tensor turns out to be a mass term in the case
of maximal symmetry. If gluons behave as massive particles
when they propagate along curved gauge slices,
it is possible to describe massiveness without mass term.
The extended gauge slices into galactic scale or inter-galactic regions may provide
new scopes for dark matter or vacuum energy.
%
%
%
%

%
%

\begin{references}


\bibitem{R1}  J. Schwinger, Phys.~Rev. {\bf 125}, 397 (1962); {\it ibid.}  {\bf 128}, 2425 (1962).
\bibitem{R2}  E. J. Eichten and F. L. Feinberg, Phys.~Rev. D~{\bf 10}, 3254 (1974).
\bibitem{R3}  J. M. Cornwall, Phys.~Rev. D~{\bf 26}, 1453 (1982).
\bibitem{R4}  M. A. Shifman, A. I. Vainshtein, and V. I. Zakharov, Nucl. Phys. B~{\bf 147}, 385 (1979);
              {\it ibid.} ~{\bf 147}, 448 (1979); {\it ibid.} ~{\bf 147}, 519 (1979).
\bibitem{R5}  T. I. Larsson, Phys.~Rev. D~{\bf 32}, 956 (1985).
\bibitem{R6}  M. Lavelle and M. Oleszczuk, Mod.~Phys.~Lett. A~{\bf 7}, 3617 (1991);
              J. Ahlbach {\it et al.}, Phys.~Lett. B~{\bf 275}, 124 (1992).
\bibitem{R7}  K. I. Kondo and T. Shinohara, Phys.~Lett. B~{\bf 491}, 263 (2000).
\bibitem{R8}  Ph. Boucaud {\it et al.}, Phys.~Lett. B~{\bf 493}, 315 (2000).
\bibitem{R9}  C. G. Callan, R. F. Dashen, and D. J. Gross, Phys.~Lett. B~{\bf 63}, 334 (1976).
\bibitem{R10} S. J. Brodsky and R. Shrock, Phys.~Lett. B~{\bf 666}, 95 (2008).
\bibitem{R11} D. Kharzeev, Phys.~Lett. B~{\bf 633}, 260 (2006).
\bibitem{R12} D. Kharzeev, R. D. Pisarski, and M. H. G. Tytgat, Phys.~Rev.~Lett.~{\bf 81}, 512 (1998).
\bibitem{R13} A. Casher and L. Susskind, Phys.~Rev. D~{\bf 9}, 436 (1974).
\bibitem{R14} S. J. Brodsky and R. Shrock, Proc.~Nat.~Acad.~Sci.~{\bf 108}, 45 (2011).
\bibitem{R15} S. A. Voloshin, Phys.~Rev. C~{\bf 70}, 057901 (2004).
\bibitem{R16} B. I. Abelev {\it et al.} (STAR Collaboration), Phys.~Rev.~Lett.~{\bf 103}, 251601 (2009);
              B. Abelev {\it et al.} (ALICE Collaboration), {\it ibid.}~{\bf 110}, 012301 (2013).
\bibitem{R17} Ph. Boucaud {\it et al.}, Phys.~Rev. D~{\bf 66}, 034504 (2002).
\bibitem{R18} E. M. Kim {\it et al.}, J.~Korean~Phys.~Soc.~{\bf 64}, 1272 (2014).
\bibitem{R19} E. J. Kim and J. B. Choi, J.~Korean~Phys.~Soc.~{\bf 61}, 1215 (2012).
\bibitem{R20} S. V. Mikhailov and A. V. Radyushkin, Phys.~Rev. D~{\bf 45}, 1754 (1992).
\bibitem{R21} E. J. Kim {\it et al.}, J.~Korean~Phys.~Soc.~{\bf 58}, 1053 (2011).
\bibitem{R22} E. J. Kim and J. B. Choi, J.~Korean~Phys.~Soc.~{\bf 64}, 495 (2014).
\bibitem{R23} S. Weinberg, {\it Gravitation and Cosmology} (John Wiley \& Sons, New York, 1972), p140.
\bibitem{R24} Ref.~\cite{R23}, p383.
\bibitem{R25} F. Sarazin {\it et al.}, Phys.~Rev. C~{\bf 70}, 031302 (2004).
\bibitem{R26} T. Ichikawa {\it et al.}, Phys.~Rev.~Lett.~{\bf 107}, 112501 (2011).
\bibitem{R27} S. K. Lee, E. J. Kim, and  J. B. Choi, arXiv:1503.08984.
\bibitem{R28} D. H. Weinberg {\it et al.}, Phys.~Rep.~{\bf 530}, 87 (2013).


\end{references}
\end{document}